\documentclass[letter,12pt]{article}

\usepackage[english]{babel}
\usepackage{booktabs}
\usepackage{rotating}
\usepackage{threeparttable, tablefootnote}
\usepackage{multirow}
\usepackage{color}
\usepackage{xcolor}
\usepackage{enumitem}
\usepackage{graphicx}
\usepackage{url} 

\usepackage{changepage}   
\usepackage[margin=1in]{geometry}
\usepackage{amsfonts}
\usepackage{amsmath}
\usepackage{amssymb}
\usepackage{bm}
\usepackage{dsfont}
\usepackage{xfrac}
\usepackage{mathtools}

\usepackage{floatrow}   
\usepackage{subcaption}

\usepackage{subfig}

\usepackage{natbib}
\renewcommand{\b}[1]{\mathbf{#1}}               
\newcommand{\norm}[1]{\left\lVert#1\right\rVert} 

\newcommand{\mc}[1]{\mathcal{#1}}

\newcommand{\normQuad}[1]{\norm{#1}_2^2}
\newcommand{\bigO}{\mathcal{O}}

\DeclareMathOperator*{\argmin}{arg\,min}

\usepackage{algorithm2e}
\SetKwComment{Comment}{/* }{ */}
\RestyleAlgo{ruled}

\usepackage{authblk}

\begin{document}

\title{An Alternating Direction Method of Multipliers Algorithm for the Weighted Fused LASSO \\ Signal Approximator}

\author[1]{Louis Dijkstra}
\author[1]{Moritz Hanke}
\author[1,2]{Niklas Koenen}
\author[1]{Ronja Foraita\footnote{Corresponding author. E-mail: \texttt{foraita@leibniz-bips.de}}}

\affil[1]{Leibniz Institute for Prevention Research \& Epidemiology -- BIPS, Bremen, Germany}

\affil[2]{Faculty of Mathematics \& Computer Science, University of Bremen, Bremen, Germany}


\maketitle

\begin{abstract}
\noindent 
We present an Alternating Direction Method of Multipliers (ADMM) algorithm designed to solve the Weighted Generalized Fused LASSO Signal Approximator (wFLSA). First, we show that wFLSAs can always be reformulated as a Generalized LASSO problem. With the availability of algorithms tailored to the Generalized LASSO, the issue appears to be, in principle, resolved. However, the computational complexity of these algorithms is high, with a time complexity of $\bigO(p^4)$ for a single iteration, where $p$ represents the number of coefficients. To overcome this limitation, we propose an ADMM algorithm specifically tailored for wFLSA-equivalent problems, significantly reducing the complexity to $\bigO(p^2)$. Our algorithm is publicly accessible through the \texttt{R}~package~\texttt{wflsa}.\\[5pt]
\end{abstract}


\section{Introduction}
\label{sec:introduction}

Several penalized regression methods have been proposed to address identifiability challenges and enforce sparsity in high-dimensional settings, where the number of variables ($p$) exceeds the number of observations ($n$). These methods impose penalties on regression coefficients, aiming to regulate model complexity and promote more robust and interpretable results. Perhaps the most prominent among these methods is the Least Absolute Shrinkage and Selection Operator (LASSO; \citet{tibshirani1996regression}), which is commonly defined as 
\begin{equation}
    \widehat{\bm{\beta}} = \argmin_{\bm{\beta} \in \mathds{R}^p} \frac{1}{2} \norm{\b{y} - \b{X}\bm{\beta}}_2^2 + \lambda_1 \norm{\bm{\beta}}_1. \qquad \text{\textit{(LASSO)}}
    \label{eq:LASSO}
\end{equation}
Here, $\bm{\beta}$ represents a $p$-dimensional vector of coefficients, $\b{y}$ stands for the $n$-dimensional response vector, $\b{X}$ is a $(n \times p)$-dimensional matrix of observations, and $\norm{\cdot}_1$ and $\norm{\cdot}_2$ represent the $\ell_1$- and $\ell_2$-norm, respectively. The term $\lambda_1\norm{\bm{\beta}}_1$ induces sparsity by `shrinking' some of the coefficients to zero. The penalty term $\lambda_1 \geq 0$ regulates the sparseness of the solution, i.e., higher values of $\lambda_1$ lead to sparser estimated regression vectors. We assume throughout that the mean of $\b{y}$ is zero. 

A plethora of variations on the LASSO exist, one of which is the \emph{Fused LASSO}, introduced by \citet{tibshirani2005sparsity}. The Fused LASSO assumes that consecutive regression coefficients (i.e., $\beta_i$ and $\beta_{i+1}$) should be more similar than those that are `further apart'. It takes advantage of this assumption by penalizing the absolute differences between consecutive regression coefficients which encourages piecewise constant or piecewise linear solutions. 
Formally, the Fused LASSO is defined as
\begin{equation}
    \widehat{\bm{\beta}} = \argmin_{\bm{\beta} \in \mathds{R}^p} \frac{1}{2} \norm{\b{y} - \b{X}\bm{\beta}}_2^2 + \lambda_1 \norm{\bm{\beta}}_1 + \lambda_2 \sum_{i = 1}^{p - 1} |\beta_{i + 1} - \beta_i|. \qquad \text{\textit{(Fused LASSO)}} 
    \label{eq:LASSO}
\end{equation}
The additional penalty term is referred to as the smoothness penalty, as it fosters `smoother' transitions from one coefficient to the next. The associated tuning parameter, $\lambda_2 \geq 0$, governs the tolerance for differences; higher values of $\lambda_2$ lead to smaller differences between consecutive coefficients.

The LASSO and Fused LASSO are commonly used in regression analysis to model the relationship between a dependent variable ($\mathbf{y}$) and a set of independent variables ($\mathbf{X}$). However, beyond their conventional use in regression, these penalized techniques are also applied to transform the original observed vector ($\mathbf{y}$) in cases where sparsity (various values being zero) and smoothness (small differences between certain values) are expected. Within this framework, the resulting transformed vector is represented by the coefficient vector $\bm{\beta}$. Later in the text, we discuss some applications in which such a transformation might be useful.

One method to achieve such a transformation is a particular case of the Fused LASSO, where the design matrix $\mathbf{X}$ is equal to the identity matrix $\mathbf{I}$. This specific instance is referred to as the \emph{1-dimensional Fused LASSO Signal Approximator} (1-dimensional FLSA; \citet{kim2009ell_1, condat2013direct}):
\begin{equation}
    \widehat{\bm{\beta}} = \underset{\bm{\beta} \in \mathbb{R}^p}{\text{argmin}} \frac{1}{2} \lVert \mathbf{y} - \bm{\beta} \rVert_2^2 + \lambda_1 \lVert \bm{\beta} \rVert_1 + \lambda_2 \sum_{i = 1}^{p - 1} |\beta_{i + 1} - \beta_i|. \quad \text{\textit{(1-dimensional FLSA)}}
    \label{eq:1DFSLA}
\end{equation}
Here, the vector $\bm{\beta} \in \mathds{R}^p$ represents the transformed $\mathbf{y} \in \mathds{R}^p$. Similar to the regression scenario, the first penalty term induces sparsity by driving certain coefficients to zero, while the second term encourages similarity between consecutive coefficients.

The 1-dimensional FLSA finds applications in various domains such as comparative genomic hybridization \citep{tibshirani2008spatial} and chromosomal microarray analysis \citep{hoefling2010path}. Higher dimensional versions of the FLSA exist as well. The 2-dimensional version is, for example, used to denoise images, where neighboring pixels are thought to be similar \citep{xu2017sequential}, see Section~\ref{sec:applicationExample}. Another important application is change point detection, which are abrupt changes of the values in the transformed regression coefficient vector. This can be useful in a variety of fields in which systematic changes in the system of interest need to be detected \citep{rojas2014change,harris2022scalable}. 
Section~\ref{sec:applicationExample} provides an example of how the wFLSA can be applied to heterogeneous image smoothing and compares it to other algorithms. 

\citet{hoefling2010path} extended the definition of the Fused LASSO by introducing an undirected graph $G = (V,E)$, where the vertices $V = \{1,2,\ldots,p\}$ correspond to the coefficients $\beta_1, \beta_2, \ldots, \beta_p$. The adjacency of a pair of coefficients is determined by the graph's edge set $E \subseteq V \times V$, where $\{i,j \} \in E$ denotes an edge between vertices $i$ and $j$ in the graph. Based on the graph, one can generalize the FLSA as follows: 
\begin{equation}
    \widehat{\bm{\beta}} = \argmin_{\bm{\beta} \in \mathds{R}^p} \frac{1}{2} \norm{\b{y} - \bm{\beta}}_2^2 + \lambda_1 \norm{\bm{\beta}}_1 + \lambda_2 \sum_{\{ i, j\} \in E} |\beta_{i} - \beta_j|. \qquad \text{\textit{(Generalized FLSA)}} 
    \label{eq:FSLA}
\end{equation}
The Generalized FLSA encourages similarity between coefficients linked by an edge in the graph. Let $\mathbf{A}$ denote the adjacency matrix of the graph $G$. The 1-dimensional FLSA, see eq.~\eqref{eq:1DFSLA}, is a specific instance of the Generalized FLSA, as it is equivalent when the adjacency matrix is defined as
\begin{equation*}
    \b{A}_{\text{1-dimensional}} = \begin{bmatrix}
0 & 1 & 0 & \dots & 0 & 0 \\
1 & 0 & 1 & \dots & 0 & 0 \\
0 & 1 & 0 & \dots & 0 & 0 \\
\vdots & \vdots & \vdots & \ddots & \vdots & \vdots \\
0 & 0 & 0 & \dots & 0 & 1 \\
0 & 0 & 0 & \dots & 1 & 0 \\
\end{bmatrix}.
\end{equation*}
Several efficient algorithms have been developed for solving the Generalized FLSA \citep{hoefling2010path,qian2016stepwise,son2019modified}.

In this context, we consider an even more general version: the \emph{Weighted Generalized Fused LASSO Signal Approximator} (wFLSA) in which the edges in the graph are weighted. Consider the graph $G = (V, E, w)$, where $w: V \times V \rightarrow [0, \infty)$ is a weight function assigning non-negative values to each possible edge $\{i,j\}$. Let the weighted adjacency matrix of the graph be denoted as $\mathbf{W} = \{w_{ij}\}$, where $w_{ij} = w(\{i,j\})$. The wFLSA is defined as 
\begin{equation}
     \widehat{\bm{\beta}} = \argmin_{\bm{\beta} \in \mathds{R}^p} \frac{1}{2} \norm{\b{y} - \bm{\beta}}_2^2 + \lambda_1 \norm{\bm{\beta}}_1 + \lambda_2 \sum_{i<j} w_{ij} |\beta_{i} - \beta_j|. \quad \text{\textit{(Weighted Generalized FLSA)}} 
    \label{eq:wFLSA}
\end{equation}
Not only does this approach allow for determining which coefficients should be similar, but it also provides the flexibility to regulate the degree of similarity by assigning varying weights to the edges. It is evident that the 1-dimensional and Generalized FLSA are special cases of this weighted version.

To the best of our knowledge, no algorithm for the wFLSA exists in the literature. Our approach to estimating a wFLSA follows two main steps. First, we establish that the wFLSA can be reformulated as a special case of the Generalized LASSO framework \citep{tibshirani2011generalized}, see Section~\ref{sec:wFLSAasGeneralizedLASSO}. Throughout, we refer to Generalized LASSO problems of this kind as \emph{wFLSA-equivalent problems}. By demonstrating their equivalence, we essentially resolve the problem at hand. Existing path algorithms for solving the Generalized LASSO, see \citet{arnold2016efficient} and \citet{zhao2020solution}, theoretically possess the capability to address problems equivalent to wFLSAs.
However, the computational complexity associated with these path algorithms becomes prohibitively high for wFLSA-equivalent problems, i.e., in the order of $\mathcal{O}(N_{\text{GL}} \cdot p^4)$, where $N_{\text{GL}}$ is the number of iterations needed for the algorithm to converge \citep{arnold2016efficient,zhao2020solution}. 
This makes it ill-suited for situations with high-dimensional data characterized by a large number of coefficients. 

To overcome this computational limitation, we proceed with the second step, in which we introduce a tailored Alternating Direction Method of Multipliers (ADMM; \citet{boyd2004convex}) algorithm designed specifically to tackle wFLSA-equivalent problems, see Section~\ref{sec:ADMM}. 
The ADMM algorithm, an optimization method used in a variety of fields  \citep{forero2010consensus,wahlberg2012admm,yang2018admm,glowinski2022application}, divides the original optimization problem into three subproblems \citep{boyd2004convex}. At each iteration, the ADMM updates 1) the primal variables, 2) the dual variables, and 3) a penalty parameter (hence its name `alternating direction').
The primal variables correspond to the variables of interest in the original optimization problem. Updating them involves minimizing an augmented Lagrangian function, consisting of components of the original objective function, alongside Lagrange multipliers terms to enforce the constraints.
Subsequently, the algorithm updates the dual variables based on the differences between the current value of the primal variables and the constraints. 
The ADMM's penalty parameter regulates the trade-off between staying close to the original problem on the one hand and adhering to the constraints on the other. 
The ADMM proves particularly advantageous for handling large-scale data, as it can be parallelized \citep{boyd2004convex}. 

The proposed ADMM algorithm notably reduces the complexity to $\mathcal{O}(N_{\text{ADMM}} \cdot p^2)$, where $N_{\text{ADMM}}$ is the number of iterations for the algorithm to converge (see Section~\ref{sec:computationalComplexity}), rendering it more scalable to real-world problems. 

The paper is structured as follows. In Section~\ref{sec:wFLSAasGeneralizedLASSO}, we demonstrate how the wFLSA can be represented as a special instance of the Generalized LASSO. We continue by describing the ADMM algorithm tailored to these wFLSA-equivalent Generalized LASSO problems in Section~\ref{sec:ADMM}. We show in Section~\ref{sec:computationalComplexity} that, by leveraging the unique structure, the computational complexity of the ADMM algorithm is $\mathcal{O}(N_{\text{ADMM}} \cdot p^2)$. 
Section~\ref{sec:tuningParameterWeightMatrixSelection} discusses various methods for choosing the weight matrix and the two tuning parameters based on data. Section~\ref{sec:applicationExample} provides an example of how the wFLSA can be used for heterogeneous image smoothing. Finally, in Section~\ref{sec:conclusionsDiscussion}, a conclusion and suggestions for future research are provided. 
Appendix~\ref{sec:appendixExampleGeneralizedLASSO} includes an example of how the wFLSA can be formulated as a Generalized LASSO problem when $p = 4$. In addition, Appendix~\ref{sec:appendixChoiceOfMatrixQADMM} provides information on how to choose matrix $Q$ used in the ADMM efficiently. An implementation of the algorithm is publicly available as an \texttt{R}~package under \url{https://github.com/bips-hb/wflsa}. The code for the example in Section~\ref{sec:applicationExample} can be found at \url{https://github.com/bips-hb/wFLSA-paper}. 

\section{The Weighted Fused LASSO Signal Approximator as a Special Case of the Generalized LASSO} \label{sec:wFLSAasGeneralizedLASSO}

First, we use a fundamental result in the work of \citet{Friedman2007}, in which they introduce a pathwise coordinate optimization algorithm used to solve the 1-dimensional FLSA, among others. Let us denote the estimate for the regression coefficients of the wFLSA for specific tuning parameters and a given weight matrix as 
\begin{equation*}
    \begin{split}
        \widehat{\bm{\beta}}(\lambda_1, \lambda_2, \mathbf{W}) & = \left[ \widehat{\beta}_1(\lambda_1,\lambda_2,\b{W}), \widehat{\beta}_2(\lambda_1,\lambda_2,\b{W}), \ldots, \widehat{\beta}_p(\lambda_1,\lambda_2,\b{W})\right]^\top \\ & =  \arg\min_{\bm{\beta} \in \mathbb{R}^p} \frac{1}{2} \|\mathbf{y} - \bm{\beta}\|_2^2 + \lambda_1 \|\bm{\beta}\|_1 + \lambda_2 \sum_{i<j} w_{ij} |\beta_{i} - \beta_j|.
    \end{split}
\end{equation*}
\citet{Friedman2007} demonstrated that once an estimate $\widehat{\bm{\beta}}(0, \lambda_2, \b{W})$ is obtained (the wFLSA without the sparsity penalty, i.e., $\lambda_1 = 0$), one can readily derive the estimate $\widehat{\bm{\beta}}(\lambda_1, \lambda_2, \mathbf{W})$ for any $\lambda_1 > 0$. This is achieved by applying a soft thresholding function to each of the values of $\widehat{\bm{\beta}}(0, \lambda_2, \mathbf{W})$. The soft thresholding function is defined as
\begin{equation}
    \text{ST}_{\lambda_1}(x) := \begin{cases}
x - \lambda_1 & \text{if } x > \lambda_1, \\ 
0 & \text{if } -\lambda_1 \leq x \leq \lambda_1, \\ 
x + \lambda_1 & \text{otherwise}. 
\end{cases}
\label{eq:definitionSoftThresholdingFunction}
\end{equation}
The estimate $\widehat{\bm{\beta}}(\lambda_1, \lambda_2, \b{W})$ is then given by
\begin{equation*}
    \widehat{\bm{\beta}}(\lambda_1, \lambda_2, \b{W}) = \left[ \text{ST}_{\lambda_1}\left(\widehat{\beta}_1(0,\lambda_2,\b{W})\right), \text{ST}_{\lambda_1}\left(\widehat{\beta}_2(0,\lambda_2,\b{W})\right), \ldots, \text{ST}_{\lambda_1}\left(\widehat{\beta}_p(0,\lambda_2,\b{W})\right) \right]^\top.
\end{equation*}
This result is important for three reasons. First, it allows us to focus solely on solving the simpler problem $\widehat{\bm{\beta}}(0,\lambda_2, \mathbf{W})$, which we will do throughout the rest of the paper. 
Second, determining the estimate for a different $\lambda_1 > 0$ is computationally efficient, as applying soft thresholding to a $p$-dimensional vector requires $\bigO(p)$ steps.
Third, and perhaps most importantly, we can identify the specific $\lambda_1$ values at which the $i$-th coefficient  enters the active set, i.e., becomes non-zero. This stems from the fact that $\text{ST}_{\lambda_1}(\widehat{\beta}_i(0,\lambda_2,\b{W}))$ only becomes non-zero if $ |\widehat{\beta}_i(0,\lambda_2,\b{W})| > \lambda_1$. Consequently, for a fixed value of $\lambda_2$, we know the entire \emph{solution path} of $\lambda_1$. This greatly simplifies the problem at hand and substantially reduces the computation time required.

Now, we show that the problem of estimating $\widehat{\bm{\beta}}(0, \lambda_2, \b{W})$ can be reformulated as a Generalized LASSO problem. The Generalized LASSO, as introduced by \citet{tibshirani2011generalized}, is formulated as follows:
\begin{equation}
    \widehat{\bm{\beta}} = \argmin_{\bm{\beta} \in \mathds{R}^p}~ \frac{1}{2} \normQuad{\b{y} - \b{X} \bm{\beta}} + \gamma \norm{\b{D} \bm{\beta}}_1. \qquad \text{\textit{(Generalized LASSO)}}
\label{eq:generalizedLASSO}
\end{equation}
Here, $\mathbf{D}$ represents a real-valued matrix with $m$ rows and $p$ columns, and $\gamma \geq 0$ serves as the tuning parameter controlling the penalty's strength. Since we are concerned with transformation rather than classic regression analysis, see Section~\ref{sec:introduction}, the design matrix $\b{X}$ is equal to  $\b{I}$. 
We demonstrate that it is possible to choose the matrix $\b{D}$ such that the sparsity penalty term from the Generalized LASSO is equal to the smoothness penalty term of the wFLSA. We do this by using the concept of the \emph{oriented incidence matrix} \citep{johnson1972graph,tibshirani2011generalized}. 

Let $\widetilde{G}$ denote a directed version of the undirected graph $G$. Specifically, let $\widetilde{w}$ represent the weight function assigning weights to each edge in the graph $\widetilde{G}$ as follows:
\begin{equation}
    \widetilde{w}(\{i,j\}) = \begin{cases}
               w(\{i,j\}) = w_{ij} & \text{ if } i < j \text{ and } \\ 
              0 & \text{otherwise}, 
    \end{cases}
    \label{eq:definitionWeightMatrixDirectedGraph}
\end{equation}
where $w$ is the weight function of the original graph $G$. In other words, the edge $\{i,j\}$ in the original graph translates to the directed edge $i \rightarrow j$ in $\widetilde{G}$ for $i < j$. We now define $\b{D} \in \mathds{R}^{\binom{p}{2} \times p}$ to be the oriented incidence matrix of the graph $\widetilde{G}$, where each row represents a potential edge $\{i,j\} \in V \times V$.  The columns correspond to the $p$ vertices/coefficients. Let $\mathbf{d}_{ij}$ denote the $p$-dimensional row vector related to the edge $\{i,j\}$, such that
\begin{equation}
    \mathbf{d}_{ij} = \Big[d_{ij}(1), d_{ij}(2), \ldots, d_{ij}(p) \Big], \qquad \text{where} \quad d_{ij}(l) = \begin{cases}
        w_{ij} & \text{if }l = i, \\ 
        -w_{ij} & \text{if }l = j, \text{ and } \\ 
        0 & \text{otherwise}.
    \end{cases} 
    \label{eq:rowOfTheIncidenceMatrix}
\end{equation}
The oriented incidence matrix $\mathbf{D}$ is then a matrix of these row vectors, one for each potential edge, i.e., 
\begin{equation}
    \mathbf{D} = \begin{bmatrix}
        \mathbf{d}_{12} \\ 
        \mathbf{d}_{13} \\ 
        \mathbf{d}_{14} \\ 
        \vdots \\ 
        \mathbf{d}_{p-2, p-1} \\ 
        \mathbf{d}_{p-1, p}
    \end{bmatrix}.
    \label{eq:definitionOrientedIncidenceMatrix}
\end{equation}
For an example demonstrating the construction of an oriented incidence matrix $\mathbf{D}$ when there are four coefficients ($p = 4$), see Appendix~\ref{sec:appendixExampleGeneralizedLASSO}. By defining the matrix $\b{D}$ as such, we obtain
\begin{equation}
    \norm{\mathbf{D} \bm{\beta}}_1 = \sum_{i < j} w_{ij} |\beta_i - \beta_j|. 
    \label{eq:goalOfHowToDefineMatrixD}
\end{equation}
Combining these results, the Generalized LASSO problem is equivalent to a wFLSA problem when the following conditions are met:
\begin{enumerate}
    \item the sparsity tuning parameter $\lambda_1$ equals $0$;
    \item the design matrix $\b{X}$ is the identity matrix $\b{I}$; 
    \item the matrix $\b{D}$ is the oriented incidence matrix of the weighted directed graph $\widetilde{G}$ associated with the original wFLSA problem, and 
    \item the tuning parameter of the Generalized LASSO $\gamma$ equals the smoothness penalty term of the wFLSA, i.e., $\gamma = \lambda_2$.  
\end{enumerate}
In conclusion, 
\begin{equation}
    \begin{split}
        \widehat{\bm{\beta}}(0, \lambda_2, \b{W}) & = \argmin_{\bm{\beta} \in \mathds{R}^p}~ \frac{1}{2} \normQuad{\b{y} -  \bm{\beta}} + \lambda_2 \sum_{i < j} w_{ij} |\beta_i - \beta_j| \\ & = \argmin_{\bm{\beta} \in \mathds{R}^p}~ \frac{1}{2} \normQuad{\b{y} -  \bm{\beta}} + \lambda_2 \norm{\b{D} \bm{\beta}}_1.
    \end{split}
\end{equation}
As mentioned earlier, showing that a wFLSA can be reformulated as a Generalized LASSO problem in principle resolves the issue, given the existence of algorithms for solving the Generalized LASSO \citep{arnold2016efficient,zhao2020solution}. Nonetheless, their computational complexity is $\bigO(N_{\text{GL}} \cdot p^4)$. To overcome this limitation, we introduce an ADMM tailored to efficiently tackle wFLSA-equivalent problems in the following section.

\section{An Alternating Direction Method of Multipliers for wFLSA-equivalent Problems} \label{sec:ADMM}

The results presented in this section rely heavily on the work done by \citet{zhu2017augmented}, wherein an ADMM tailored for the Generalized LASSO is developed. In an ADMM algorithm, the original problem is decomposed into three smaller, more manageable steps, see Section~\ref{sec:introduction}. However, \citet{zhu2017augmented} demonstrates that in the context of the Generalized LASSO, the algorithm can be simplified, reducing the number of update steps to two. We refer to these two steps as the $\bm{\beta}$- and $\bm{\alpha}$-update step. 

For each update step indexed by $k = 0,1,2,\ldots$, we consider two vectors: $\bm{\beta}(k)$ and $\bm{\alpha}(k)$. The former is a $p$-dimensional vector, while the latter is a vector whose dimensionality equals the number of rows of matrix $\b{D}$, i.e., $\binom{p}{2}$.

\paragraph{The $\bm{\beta}$-update step:}%
The $\bm{\beta}$-update step, as proposed by \citet{zhu2017augmented}, is expressed as
\begin{equation*}
    \bm{\beta}(k + 1) = \big(\rho \b{Q} + \b{I}\big)^{-1} \Big[ \rho \b{Q} \bm{\beta}(k) + \b{y} - \lambda_2 \cdot \b{D}^\top \big( 2 \bm{\alpha}(k) - \bm{\alpha}( k -  1) \big) \Big] \quad \text{for } k = 1,2,\ldots.
\end{equation*}
Here, $\b{Q}$ denotes a $(p \times p)$-dimensional matrix, $\rho > 0$ stands for the ADMM's dual update step length \citep{boyd2004convex,zhu2017augmented}, $\b{y}$ is the $p$-dimensional vector of observations, $\lambda_2$ is the smoothness penalty term and $\b{D}$ is the oriented incidence matrix corresponding to the wFLSA problem under consideration. 

The choice of matrix $\mathbf{Q}$ plays a critical role, as it can significantly accelerate the algorithm when chosen appropriately \citep{zhu2017augmented}. It must satisfy the condition that the matrix $\mathbf{Q} - \lambda_2^2 \cdot \mathbf{D}^\top \mathbf{D} \succ 0$ is positive definite. For simplicity, we choose $\mathbf{Q} = q\mathbf{I}$, i.e., a matrix with the constant $q > 0$ along the diagonal. To ensure positive definiteness, $q$ must exceed the maximum eigenvalue of $\lambda_2^2 \cdot \mathbf{D}^\top \mathbf{D}$, which can be computed numerically. However, performing the matrix multiplication $\mathbf{D}^\top\mathbf{D}$ can be challenging due to the potentially large number of rows in $\mathbf{D}$. For instance, with $p = 1{,}000$, the row count of $\mathbf{D}$ is close to half a million; for $p = 10{,}000$, it approaches 50 million, and so on. Fortunately, $\mathbf{D}^\top \mathbf{D}$ possesses a distinct structure that substantially reduces the computational time required (see Appendix~\ref{sec:appendixChoiceOfMatrixQADMM}). By defining $\mathbf{Q}$ as $q\mathbf{I}$, the update step for $\bm{\beta}$ simplifies to
\begin{equation}
    \bm{\beta}(k + 1) = \frac{1}{\rho q + 1} \Big[ \rho q \bm{\beta}(k) + \b{y} - \lambda_2 \cdot \b{D}^\top \big( 2 \bm{\alpha}(k) - \bm{\alpha}( k -  1) \big) \Big] \quad \text{for } k = 1,2,\ldots
    \label{eq:betaUpdateStep}
\end{equation}

\paragraph{The $\bm{\alpha}$-update step:}%
Following the $\bm{\beta}$ update, we proceed by updating the value of $\bm{\alpha}$. As in the work by \citet{zhu2017augmented}, the $\bm{\alpha}$-update step is expressed as:
\begin{equation}
    \bm{\alpha}(k + 1) = \text{T}\big( \bm{\alpha}(k) + \rho \cdot \lambda_2 \cdot \b{D} \bm{\beta}(k+1)\big).
    \label{eq:alphaUpdateStep}
\end{equation}
Here, $\rho$ is again the dual update step length. The function $\text{T}(\cdot): \mathds{R}^{\binom{p}{2}}\rightarrow [-1,1]^{\binom{p}{2}}$ denotes a threshold function defined as
\begin{equation}
   \text{T}\big(x_1, x_2, \ldots, x_m\big) = \big( \widetilde{x}_1, \widetilde{x}_2, \ldots, \widetilde{x}_m  \big)^\top 
   \label{eq:thresholdingFunction}
\end{equation}
where $\widetilde{x}_i$ is determined as 
$$
\widetilde{x}_i = \begin{cases}
        -1 & \text{if } x_i < -1, \\ 
        1 & \text{if }x_i > 1 \text{ and } \\
        x_i & \text{otherwise}, 
    \end{cases}
$$
for $i = 1,2,\ldots,m$. This function restricts each component of the input vector to the range $[-1, 1]$, setting values below $-1$ to $-1$, values above $1$ to $1$, and leaving values within the range unchanged.
We iterate through the update steps for $\bm{\beta}$ and $\bm{\alpha}$ until a stopping condition is met. The algorithm terminates when the sum of the absolute differences between $\bm{\beta}(k + 1)$ and $\bm{\beta}(k)$ falls below a predefined threshold $\epsilon > 0$:
\begin{equation}
\norm{\bm{\beta}(k+1) - \bm{\beta}(k)}_1 < \epsilon.
\label{eq:stoppingCondition}
\end{equation}
Combining the findings from this section with those from the previous sections, we arrive at the following pseudo-algorithm: 
\begin{enumerate}
    \item Let $\b{y} \in \mathds{R}^p$ be the vector containing the observations, and let $G = (V, E, w)$ denote the undirected graph associated with the wFLSA, characterized by the weighted adjacency matrix $\b{W}$. It is necessary to select the tuning parameters $\lambda_2 \geq 0$, the dual update step length $\rho > 0$ (by default $\rho = 1$), and $\epsilon > 0$, which is the convergence threshold for the algorithm; 
    \item Determine the directed graph $\widetilde{G}$, as per eq.~\eqref{eq:definitionWeightMatrixDirectedGraph}, along with its oriented incidence matrix $\b{D}$ as defined in eq.~\eqref{eq:definitionOrientedIncidenceMatrix};
    \item Compute numerically the largest eigenvalue $\kappa_{\text{max}}$ (or an upper bound) of the matrix $\lambda_2^2 \cdot \mathbf{D}^\top \mathbf{D}$. See Appendix~\ref{sec:appendixChoiceOfMatrixQADMM} for further details. Select $q$ to be greater than the largest eigenvalue, i.e., $q > \kappa_{\text{max}}$. This choice guarantees that $q\mathbf{I} - \lambda_2^2 \cdot \mathbf{D}^\top \mathbf{D}$ is strictly positive definite;
    \item Initialize the vectors $\bm{\beta}(0) = \bm{\beta}(1) = \b{0}$ and $\bm{\alpha}(0) = \bm{\alpha}(1) = \b{0}$. 
    \item Execute the $\bm{\beta}$-update step according to eq.~\eqref{eq:betaUpdateStep};
    \item Execute the $\bm{\alpha}$-update step according to eq.~\eqref{eq:alphaUpdateStep}, and 
    \item If $\norm{\bm{\beta}(k+1) - \bm{\beta}(k)}_1 < \epsilon$, terminate the algorithm and return $\bm{\beta}(k+1)$ as the coefficient vector $\widehat{\bm{\beta}}(0, \lambda_2, \b{W})$. Otherwise, repeat steps 5 and 6.
\end{enumerate}
The estimates $\widehat{\bm{\beta}}(\lambda_1, \lambda_2, \b{W})$ with $\lambda_1 > 0$ can be easily obtained by applying soft thresholding, see eq.~\eqref{eq:definitionSoftThresholdingFunction}. 
An implementation of this algorithm is available  at \\ \url{https://github.com/bips-hb/wflsa}.

\section{Computational Complexity} \label{sec:computationalComplexity}

To assess the computational complexity of ADMM, we analyze the update steps for both $\bm{\beta}$ and $\bm{\alpha}$, separately. 

\subsection{The $\bm{\beta}$-Update Step}

The most computationally intensive aspect of the $\bm{\beta}$-update step, see eq.~\eqref{eq:betaUpdateStep}, is the matrix-vector multiplication
$$
\b{D}^\top \big( 2 \bm{\alpha}(k) - \bm{\alpha}( k -  1) \big).
$$
Here, $\b{D}^\top$ has $p$ rows and $\binom{p}{2}$ columns, and $\bm{\alpha}(k)$ and $\bm{\alpha}(k-1)$ are $\binom{p}{2}$-dimensional vectors. A naive approach would entail multiplying two $\binom{p}{2}$-dimensional vectors $p$ times (once for every row in $\b{D}^\top$), resulting in a complexity of $\bigO(p^3)$.

However, we can show that each row of the transposed oriented incidence matrix $\b{D}^\top$ contains at most $p-1$ non-zero entries. Hence, instead of multiplying $\binom{p}{2}$ values, only $p-1$ multiplications are needed. Given that this operation must be performed for each of the $p$ rows, the complexity is thereby reduced to $\bigO(p^2)$.

To confirm this assertion, let us start by denoting the $l$-th column vector of matrix $\b{D}$ by $\b{e}_l \in \mathds{R}^{\binom{p}{2}}$:
\begin{equation}
    \b{e}_l^\top = \Big[ d_{12}(l), d_{13}(l), d_{14}(l), \ldots, d_{p-2,p}(l), d_{p-1,p}(l) \Big]^\top \quad \text{for }l = 1,2,\ldots,p.
    \label{eq:definitionColumnIncidenceMatrixE}
\end{equation}
By using the definition of $d_{ij}(l)$ from eq.~\eqref{eq:rowOfTheIncidenceMatrix}, it is evident that at most $p - 1$ entries can be non-zero (an entry is non-zero only if the corresponding entry in the weight matrix is non-zero as well). This observation leads us to conclude that each row of $\b{D}^\top$ has at most $p-1$ non-zero entries. Consequently, the expression $\b{e}_l^\top ( 2 \bm{\alpha}(k) - \bm{\alpha}( k - 1) )$ involves at most $p-1$ multiplications. Given there are $p$ rows, the complexity of the $\bm{\beta}$-update step is indeed~$\bigO(p^2)$.

\subsection{The $\bm{\alpha}$-Update Step}

The most computationally intensive aspect of the $\bm{\alpha}$-update step, see eq.~\eqref{eq:alphaUpdateStep}, involves the matrix-vector multiplication $\b{D}\bm{\beta}(k+1)$. If approached naively, this operation would yield a complexity of $\bigO(p^3)$.
The structural properties of the matrix $\b{D}$ can be utilized in this context as well. By employing the definition of $\b{D}$  in eq.~\eqref{eq:definitionOrientedIncidenceMatrix}, we can express
$$
 \b{D} \bm{\beta}(k+1) = 
    \begin{bmatrix}
    \b{d}_{12} \cdot \bm{\beta}(k+1)\\ 
    \b{d}_{13} \cdot \bm{\beta}(k+1)\\
     \b{d}_{14} \cdot \bm{\beta}(k+1)\\ 
     \vdots \\
     \b{d}_{p-2,p} \cdot \bm{\beta}(k+1)\\ 
     \b{d}_{p-1,p} \cdot \bm{\beta}(k+1)
    \end{bmatrix} 
$$
where 
$$
    \b{d}_{ij} \cdot \bm{\beta}(k+1) = w_{ij} \beta_i(k+1) - w_{ij} \beta_j(k+1) = w_{ij} \left( \beta_i(k + 1) - \beta_j(k + 1)\right)
$$
for all $i < j$. Each entry of the vector $\b{D} \bm{\beta}(k + 1)$ thus requires a constant number of operations, specifically a summation and a multiplication. Since the vector has $\binom{p}{2}$ entries, the complexity is of the order $\bigO(p^2)$. Consequently, computing $\b{D}\bm{\beta}(k+1)$ requires $\bigO(p^2)$ computations. 

As mentioned earlier, selecting an appropriate value for $q > 0$ is crucial to ensure that the matrix $\lambda_2^2 \cdot  \mathbf{D}^\top \mathbf{D}$ is positive definite. Computing the largest eigenvalue of this matrix numerically requires $\bigO(p^3)$ operations, which is manageable for small values of $p$. However, as $p$ increases, this computational cost can become prohibitive. Fortunately, it suffices to establish an upper bound for the largest eigenvalue, which can be achieved using the Gershgorin circle theorem \citep{gloub1996matrix}. Importantly, this process can be accomplished with a complexity of $\bigO(p^2)$. Further information is provided in Appendix~\ref{sec:appendixChoiceOfMatrixQADMM}.

In conclusion, given that both the $\bm{\beta}$- and $\bm{\alpha}$-update steps have a complexity of $\bigO(p^2)$, a complete single update step has a complexity of $\bigO(p^2)$ as well. Let $N_\text{ADMM}$ be the number of iterations required to satisfy the stopping criterion defined in eq.~\eqref{eq:stoppingCondition}, the overall computational complexity of the ADMM algorithm is $\bigO(N_\text{ADMM} \cdot p^2)$.

\section{Tuning Parameter and Weight Matrix Selection}\label{sec:tuningParameterWeightMatrixSelection}

The above-described ADMM algorithm requires the selection of appropriate values for:
\begin{enumerate}
    \item the tuning parameter $\lambda_1$, which controls the sparsity of the coefficient vector; 
    \item the tuning parameter $\lambda_2$, which manages the level of smoothness or similarity among different coefficients, and 
    \item a weight matrix $\mathbf{W}$, determining which coefficients are to be similar and to what degree.
\end{enumerate} Choosing suitable values for these parameters is challenging, and only a few research papers have addressed this issue. For instance, \citet{Gao2021} proposed using the `classical' approaches AIC, BIC, and General Cross Validation (GCV) for their Least Absolute Derivations Fused LASSO Signal Approximator, but did not provide a comprehensive comparison. \citet{son2019modified} compared the BIC and the extended BIC (eBIC) for the FLSA, where the latter imposes a stronger penalty on larger models. As a result, eBIC achieved a better false discovery rate with only a slightly lower true positive rate.
Recently, \citet{Son2023} proposed a two-step selection procedure based on a General Information Criterion (GIC). In the first step, GIC is used to select $\lambda_1$, and in the second step, a preliminary test statistic with a piece-wise mean structure is used to select $\lambda_2$. Their numerical results show a good false discovery rate under the assumption of low (Gaussian) noise.
For the wFLSA, we also need to search over a finite number of weight matrices in $\mathbf{W} \in \mathcal{W}$. This makes approaches such as cross-validation or stability selection quite challenging due to their subsampling related algorithms. Moreover, even for information criteria such as eBIC, the computational cost seems to be quiet high.

\section{Heterogeneous Image Smoothing} \label{sec:applicationExample}

In addition to its originally intended applications for regression problems, the smoothing property can also be utilized for image processing. Generally, LASSO or Fused LASSO have already been employed for sparse coding or image denoising \citep{Li2014,Agarwal2007}. However, these methods, along with many traditional Bayesian approaches, often assume the presence of additive zero-mean and homogeneous Gaussian noise in the image \citep{Elad2023}. This means the noise intensity is uniform across all pixels, commonly referred to as additive white Gaussian noise. In reality, images often exhibit pixel-dependent heterogeneous noise structures, which also pose challenges for the deep learning methods that have recently become prevalent \citep{Zhou2019WhenAD}.

By using the weight matrix $\mathbf{W}$, prior knowledge can be incorporated to apply the smoothing effect only to certain regions rather than uniformly throughout the image. This approach prevents the blurring of completely clear regions in the presence of heterogeneous or spatially variant noise, a common issue with methods like the median filter. However, applying wFLSA to high-resolution images is not feasible, as the algorithm's complexity grows quadratically with the number of pixels (see Section~\ref{sec:computationalComplexity}). Given the plausible assumption of locality that pixels only interact with their local neighbors, this problem can be reduced to a patch-wise evaluation, which is a common technique in image processing, especially for deep learning-based denoisers. This technique divides an image into smaller sub-images (called patches) of a predefined size (e.g., $5 \times 5$) and applies an operation (usually called a filter or kernel) to each patch. By sliding the patch frame across the image, e.g., by moving it one pixel to the right, the results can be averaged for the overlapping pixels, providing an additional smoothing effect \citep{Szeliski2011}. In this context, wFLSA is applied to all image patches with the corresponding weight patches of $\mathbf{W}$ and then convoluted over the entire high-dimensional image.

\begin{figure}[!t]
    \centering
    \begin{subfigure}[b]{0.3333\textwidth}
         \centering
         \includegraphics[width=\textwidth]{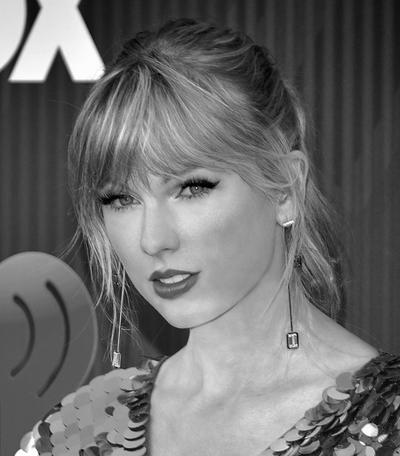}
         \caption{Original$^\text{1}$}
     \end{subfigure}%
     \begin{subfigure}[b]{0.3333\textwidth}
         \centering
         \includegraphics[width=\textwidth]{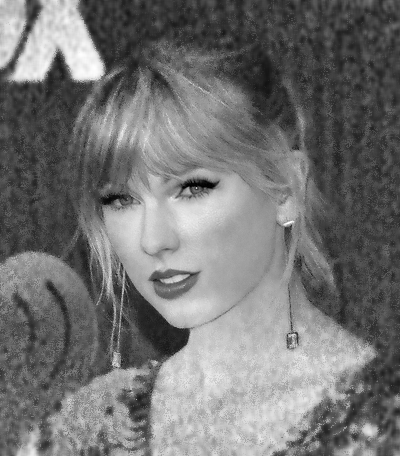}
         \caption{wFLSA ($\lambda_2 = .04$)}
     \end{subfigure}%
     \begin{subfigure}[b]{0.3333\textwidth}
         \centering
         \includegraphics[width=\textwidth]{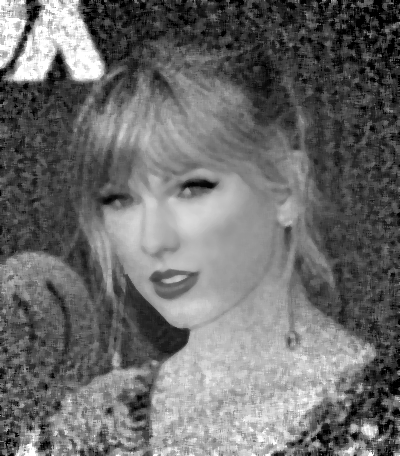}
         \caption{Median filter}
     \end{subfigure}
     \vskip5pt
     \begin{subfigure}[b]{0.3333\textwidth}
         \centering
         \includegraphics[width=\textwidth]{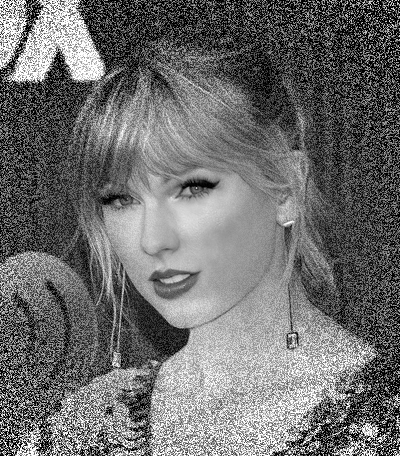}
         \caption{Radial noise}
     \end{subfigure}%
     \begin{subfigure}[b]{0.3333\textwidth}
         \centering
         \includegraphics[width=\textwidth]{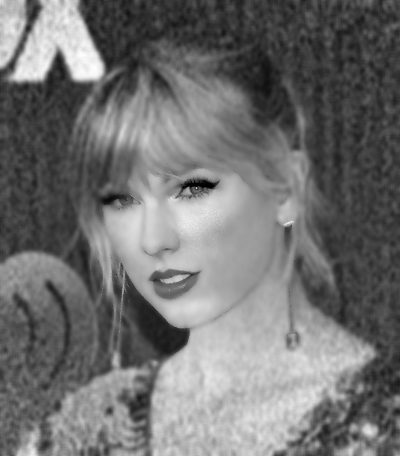}
         \caption{wFLSA ($\lambda_2 = .1$)}
     \end{subfigure}%
     \begin{subfigure}[b]{0.3333\textwidth}
         \centering
         \includegraphics[width=\textwidth]{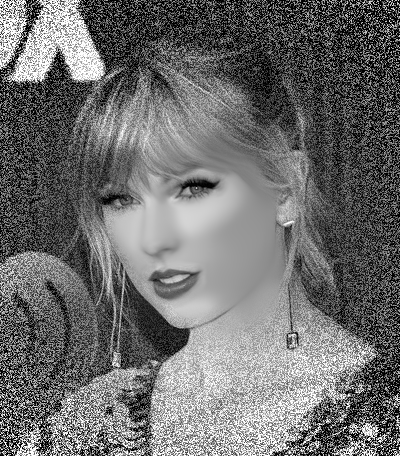}
         \caption{NLM}
     \end{subfigure}
        \caption{Application of wFLSA for image smoothing with prior knowledge on the noise structure. Image (a) is the original (a photo of Taylor Swift), and (d) is the same image with added radial noise. Images (b) and (e) are the reconstructions using wFLSA with different smoothing parameters $\lambda_2$. Image (c) uses a $5 \times 5$ median filter, and (f) employs the non-local means (NLM) method.}
        \label{fig:image_smoothing}
\end{figure}

\footnotetext[1]{\scriptsize Image source: \url{https://de.m.wikipedia.org/wiki/Datei:Taylor_Swift_2_-_2019_by_Glenn_Francis_(cropped).jpg}}

To illustrate this property, a grayscale image of size $456 \times 400$ is taken, and additive noise is added, which increases radially from the center towards the corners, resulting in higher noise levels at the edges. This intensity matrix is then used as the weight matrix $\mathbf{W}$ to apply a stronger smoothing effect in the corners compared to the central region of the image. A patch size of $5 \times 4$ is chosen for the convolution. Additionally, the denoised results are visually compared with those obtained using the median filter and the non-local means (NLM) denoising algorithm \citep{BuadesNLM}. The tuning parameter $\lambda_2$ was set to $.04$ and $.1$, varying the level of smoothness. The sparsity parameter $\lambda_1$ was fixed to $.001$ in both cases. The results are presented in Figure~\ref{fig:image_smoothing}, showing the trade-off between reconstructing and smoothing the image. The figure can be reproduced using the code available at \url{https://github.com/bips-hb/wFLSA-paper}.

\section{Conclusions and Discussion} \label{sec:conclusionsDiscussion}

We proposed an ADMM algorithm tailored for the Weighted Generalized Fused Lasso Signal Approximator (wFLSA). Our approach involved two key steps. First, we have shown that a wFLSA problem could be reformulated as a Generalized LASSO by leveraging the oriented incidence matrix of the directed version of the underlying graph of the wFLSA, see Section~\ref{sec:wFLSAasGeneralizedLASSO}. This step theoretically resolves the issue, as path algorithms exists for solving the Generalized LASSO \citep{arnold2016efficient,zhao2020solution}. However, the computational complexity of these algorithms is prohibitively high, i.e., $ \mathcal{O}(N_{\text{GL}} \cdot p^4)$ where $N_\text{GL}$ is the number of iterations needed for the algorithm to converge, which makes real-world applications challenging. To address this, we developed an ADMM approach, relying on the groundwork laid by \citet{zhu2017augmented}. We established that our ADMM-based solution drastically reduces complexity to $\mathcal{O}(N_{\text{ADMM}} \cdot p^2)$, see Section~\ref{sec:computationalComplexity}.

We discuss various methodologies for selecting the model's tuning parameters and weight matrix in Section~\ref{sec:tuningParameterWeightMatrixSelection}. This is far from trivial since the computational complexity of the ADMM algorithm makes a broad search space rather challenging. 
Here, we propose a different approach that combines a heuristic method with the use of an information criterion. Specifically, we first select a set of `reasonable' weight matrices ($\mc{W}$), possibly based on prior knowledge about a network structure. Next, for each weight matrix $\mathbf{W} \in \mc{W}$, a $\lambda_1$-value is chosen using the General Information Criterion (GIC). Finally, based on the $\mathbf{W}$ and $\lambda_1$-value that yield the smallest GIC, $\lambda_2$ is determined using a preliminary test statistic as suggested by \citet{Son2023}. It would be interesting for future research to compare this approach with the existing methodologies discussed in Section~\ref{sec:tuningParameterWeightMatrixSelection}.

In Section~\ref{sec:applicationExample}, we demonstrated how wFLSA can efficiently be used for spatially varying noise reduction by using prior knowledge on the noise intensities in the weight matrix. Our approach relaxes the very common assumption of white additive Gaussian noise and does not require an extensive training procedure, as for many deep learning techniques.

\citet{hoefling2010path} suggested the possibility of extending his path algorithm designed for solving the Generalized FLSA to the weighted version. However, how to adapt the algorithm remains unclear. 

Considering the wFLSA problem as a Generalized LASSO problem offers a particular advantage. The algorithms proposed by \citet{arnold2016efficient,zhao2020solution} are \emph{path algorithms}, providing the entire solution path for the smoothness penalty term $\lambda_2$, i.e., $\widehat{\bm{\beta}}(0, \lambda_2, \b{W})$ for all $\lambda_2 \geq 0$. This is not the case for the ADMM presented in this paper; the value of $\lambda_2$ is fixed and chosen beforehand. The proof of \citet{Friedman2007}, as discussed in Section~\ref{sec:wFLSAasGeneralizedLASSO}, shows that once the estimate of the coefficients for a specific value of $\lambda_2$ is known, the solution path for $\lambda_1 \geq 0$ is known as well. Combining these results means that when employing the Generalized LASSO, one can obtain the entire \emph{solution plane} (!), encompassing solutions for all possible values of $\lambda_1$ and $\lambda_2$. However, due to the computational complexity associated with the Generalized LASSO, using this approach may not always be feasible. 

The implementation of the application example from Section~\ref{sec:applicationExample} can be found at \url{https://github.com/bips-hb/wFLSA-paper}. The ADMM algorithm has been implemented as an \texttt{R}~package and is publicly available under \url{https://github.com/bips-hb/wflsa}.



\section*{Acknowledgments}
We gratefully acknowledge the financial support of the German Research Foundation (DFG -- Project FO 1045/2-1).

\appendix

\section{An Example: wFLSA as Generalized LASSO with $p = 4$} \label{sec:appendixExampleGeneralizedLASSO}


In Section~\ref{sec:wFLSAasGeneralizedLASSO}, we demonstrate that the wFLSA problem in eq.~\eqref{eq:wFLSA} can be reformulated as the Generalized LASSO problem as defined in eq.~\eqref{eq:generalizedLASSO}. Let us illustrate how a wFLSA problem can be translated into an equivalent Generalized LASSO problem when the number of coefficients is $p = 4$. We argue that the penalty terms of the wFLSA can be expressed as
$$
     \lambda_2 \sum_{i = 1}^3 \sum_{j = i + 1}^4 w_{ij} | \beta_i - \beta_j | = \lambda_2 \norm{\b{D} \bm{\beta}}_1.
$$ 
Here, $\lambda_2$ represent the smoothness penalty term, $\bm{\beta} = (\beta_1, \beta_2, \beta_3, \beta_4)^\top$ denotes the vector of coefficients, $w_{ij} \geq 0$ indicates the weight of the edge between the $i$-th and $j$-th node in the graph $G$, and $\b{D}$ represents the oriented incidence matrix. By using eq.~\eqref{eq:definitionOrientedIncidenceMatrix}, we find that the oriented incidence matrix equals
\begin{equation*}
    \b{D} = \begin{bmatrix}
     w_{12} & -w_{12} & 0 & 0 \\ 
     w_{13} & 0 & -w_{13} & 0 \\ 
     w_{14} & 0 & 0 & -w_{14} \\ 
     0 & w_{23} & -w_{23} & 0 \\ 
     0 & w_{24} & 0 & -w_{24} \\ 
     0 & 0 & w_{34} & -w_{34} 
    \end{bmatrix}.
\end{equation*}
Expanding the expression $\b{D}\bm{\beta}$, we observe that
\begin{equation*}
    \b{D} \bm{\beta} = \left[ \begin{array}{c c c c}
    w_{12} & -w_{12} & 0 & 0 \\ 
    w_{13} & 0 & -w_{13} & 0 \\ 
    w_{14} & 0 & 0 & -w_{14} \\ 
    0 & w_{23} & -w_{23} & 0 \\ 
    0 & w_{24} & 0 & -w_{24} \\ 
    0 & 0 & w_{34} & -w_{34} \\
\end{array}\right]
    \begin{bmatrix}
    \beta_1 \\ 
    \beta_2 \\
    \beta_3 \\
    \beta_4
    \end{bmatrix} = 
    \left[ \begin{array}{c}
    w_{12} (\beta_1 - \beta_2) \\ 
    w_{13} (\beta_1 - \beta_3) \\ 
    w_{14} (\beta_1 - \beta_4) \\ 
    w_{23} (\beta_2 - \beta_3) \\ 
    w_{24} (\beta_2 - \beta_4) \\ 
    w_{34} (\beta_3 - \beta_4) 
    \end{array} \right].
\end{equation*}
The $\ell_1$-norm of $\b{D}\bm{\beta}$ is the summation of the absolute values of the entries of the vector on the right-hand side, which yields
\begin{align*}
\lambda_2 \sum_{i < j} |w_{ij} (\beta_i - \beta_j)| = \lambda_2 \sum_{i < j} w_{ij} | \beta_i - \beta_j |.
\end{align*}
This equality holds as the weights are non-negative values.

\section{Choice of Matrix $\b{Q}$ for the ADMM Algorithm} \label{sec:appendixChoiceOfMatrixQADMM}

The ADMM algorithm introduced in this paper requires selecting a $(p \times p)$-dimensional matrix $\mathbf{Q}$ such that $\mathbf{Q} - \lambda_2^2 \cdot \mathbf{D}^\top \mathbf{D}$ is positive definite, see Section~\ref{sec:ADMM} and, more specifically, \citet{zhu2017augmented}. To make the problem more tractable, we define $\mathbf{Q}$ as a diagonal matrix with a constant $q > 0$ on its diagonal, i.e., $\mathbf{Q} = q\mathbf{I}$. Determining suitable values of $q$ can be done numerically;  if $\kappa_\text{max}$ is the largest eigenvalue of $\lambda_2^2 \cdot \mathbf{D}^\top\mathbf{D}$, then any $q > \kappa_{\text{max}}$ satisfies the condition. However, computing the product $\mathbf{D}^\top\mathbf{D}$ can be challenging due to the potentially large number of rows of $\mathbf{D}$. Fortunately, $\mathbf{D}^\top \mathbf{D}$ exhibits a distinct structure.

Recall that we can express the incidence matrix $\mathbf{D} = [\mathbf{e}_1, \mathbf{e}_2, \ldots, \mathbf{e}_p]$ in terms of its $p$ column vectors, see eq.~\eqref{eq:definitionColumnIncidenceMatrixE},  
\begin{equation*}
    \b{e}_l = \begin{bmatrix}
        d_{12}(l) \\ 
        d_{13}(l) \\ 
        d_{14}(l) \\ 
        \vdots \\ 
        d_{p-2,p}(l) \\ 
        d_{p-1,p}(l) 
    \end{bmatrix}
    \qquad \text{for }l = 1,2,\ldots,p. 
\end{equation*}
By doing so, the premultiplication of $\b{D}$ by its transpose can be written as  
\begin{equation*}
   \b{D}^\top \b{D} = \begin{bmatrix}
       \b{e}_1^\top\b{e}_1 & \b{e}_1^\top\b{e}_2 & \ldots & \b{e}_1^\top \b{e}_p  \\
       \b{e}_2^\top\b{e}_1 & \b{e}_2^\top\b{e}_2 & \ldots & \b{e}_2^\top \b{e}_p  \\
       \vdots & \vdots & \ddots & \vdots \\ 
       \b{e}_p^\top\b{e}_1 & \b{e}_p^\top\b{e}_2 & \ldots & \b{e}_p^\top \b{e}_p 
   \end{bmatrix}. 
\end{equation*}
Let us start by examining the diagonal elements. For $l = 1,2,\ldots,p$, the diagonal entry is given by
$$ 
\bar{\mathbf{w}}_l = \mathbf{e}_l^\top \mathbf{e}_l = \sum_{i < j} d_{ij}(l) d_{ij}(l) = \sum_{i < j} d_{ij}^2(l). 
$$
Using the definition of $d_{ij}(l)$ from eq.~\eqref{eq:rowOfTheIncidenceMatrix}, we deduce that
\begin{equation*}
    d_{ij}^2(l) = 
    \begin{cases}
        w_{il}^2 & \text{if }l = j, \\ 
        w_{lj}^2 & \text{if }l = i, \text{ and} \\ 
        0 & \text{otherwise}.
\end{cases} 
\end{equation*} 
This leads to a simpler expression for the diagonal value:
\begin{equation*}
    \bar{\mathbf{w}}_l = \mathbf{e}_l^\top \mathbf{e}_l = \sum_{i = 1}^{l - 1} w_{il}^2 + \sum_{j = l + 1}^p w_{lj}^2.
\end{equation*}
Due to the symmetry of the weight matrix ($w_{ij} = w_{ji}$) and its zero diagonal ($w_{ll} = 0$), we can express this as
\begin{equation*}
    \bar{\mathbf{w}}_l = \mathbf{e}_l^\top \mathbf{e}_l = \sum_{i = 1}^p w_{il}^2.
\end{equation*} 
In other words, the $l$-th entry on the diagonal of the matrix $\mathbf{D}^\top \mathbf{D}$ equals the sum of squared values of the $l$-th column (or row) of the weighted adjacency matrix $\mathbf{W}$.

The off-diagonal values are rather straightforward. For $s \neq t$, we have:
\begin{equation*}
    \mathbf{e}_s^\top \mathbf{e}_t = \sum_{i < j} d_{ij}(s) d_{ij}(t) = w_{st} \cdot (- w_{ts}) = - w_{st}^2.
\end{equation*} 
Combining these results, the matrix $\mathbf{D}^\top \mathbf{D}$ can be expressed as
\begin{equation*}
    \mathbf{D}^\top \mathbf{D} = \text{diag}(\bar{\mathbf{w}}_1, \bar{\mathbf{w}}_2, \ldots, \bar{\mathbf{w}}_p) - \mathbf{W} \circ \mathbf{W},
\end{equation*} 
where $\circ$ denotes element-wise multiplication. Numerically determining this matrix is computationally much more efficient than directly premultiplying $\b{D}$ by $\b{D}^\top$.

Consider that computing the largest eigenvalue of a $(p \times p)$-dimensional matrix is computationally intensive, scaling as $\bigO(p^3)$. While this poses no significant challenge for smaller values of $p$, it becomes a notable bottleneck for larger dimensions. Instead of directly estimating the largest eigenvalue, we can also employ an upper bound.

According to the Gershgorin circle theorem \citep{gloub1996matrix}, the largest eigenvalue is upper bounded by the maximum absolute row sum of the matrix $\mathbf{D}^\top\mathbf{D}$. 
Using our previous result, we find that the sum of the absolute values of the $i$-th row is simply $2 \bar{\mathbf{w}}_i$. An upper bound for the largest eigenvalue is, therefore, 
$$
\kappa_{\text{max}} \leq 2 \cdot \text{max} \{\bar{\mathbf{w}}_i \text{ for }i=1,2,\ldots,p\}.
$$
The computational complexity involved in determining this upper bound is of the order~$\bigO(p^2)$. 

\bibliography{main}

\end{document}